\documentclass[onecolumn, superscriptaddress, amsmath, amssymb, aps, pra]{revtex4-1}
\usepackage{graphicx}
\usepackage{dcolumn}
\usepackage{bm}
\usepackage[dvipdfm,colorlinks,linkcolor=blue, urlcolor=blue, anchorcolor=blue, citecolor=blue]{hyperref}
\usepackage{color}
\usepackage{ulem}
\usepackage{amsmath,amsfonts,amssymb}
\usepackage{booktabs}
\usepackage{float}
\usepackage{graphicx}
\usepackage{epstopdf}

\begin{document}
\title{Highly Efficient Creation and Detection of Deeply-bound Molecules via Invariant-based Inverse Engineering with Feasible Modified Drivings}
\author{Jiahui Zhang}
\email[]{y10220159@mail.ecust.edu.cn}
\affiliation{School of Physics, East China University of Science and Technology, Shanghai 200237, China}

\begin{abstract}
Stimulated Raman Adiabatic Passage (STIRAP) and its variants, such as multi-state chainwise-STIRAP allow efficiently transferring the populations in multi-state system and have been widely used to prepare ultracold deeply-bound molecules. However, their transfer efficiencies are generally imperfect. The main obstacle is the presence of losses and the requirement to make the dynamics adiabatic. To this end, in the present paper a theoretical method for the efficient and robust creation and detection of deeply-bound molecules is proposed. The simple three- and five-level systems with states chainwise coupled by optical fields are considered. In the regime of large detuning, the major molecular losses are pre-suppressed by reducing the dynamics of the three- and five-level molecular systems to those of effective two- and three-level counterparts, respectively. Consequently, two-level counterpart can be directly compatible with two kinds of ``Invariant-based Inverse Engineering" (IIE) recipes, the results show that both protocols give comparable performance and have good experimental feasibility. For the five-level case, by considering a relation among the four incident pulses, we show that the M-type structure can be generalized into an effective $\Lambda$-type one with the simplest resonant coupling. Therefore, this generalized model can also be directly compatible with ``IIE" recipe. Numerical calculations show that the weakly-bound molecules can be efficiently transferred to their deeply-bound states without strong laser intensity, and the stability against parameter variations is well preserved. Finally, the detection of ultracold deeply-bound molecules is discussed, the results show that all the protocols allow efficient detection of molecules.
\end{abstract}

\maketitle

\section{\label{sec:level1}Introduction}
Ultracold molecular gases are considered prime candidates for the study of many branches of fundamental physics and chemistry \cite{10.1063/1.3357286, Q2012, D1CS01040A, Dulieu_2009, B802322K}. The potential applications range from ultracold chemistry \cite{10.1063/1.4964096},  precision measurements \cite{Ulmanis2012} to quantum computation \cite{PhysRevLett.88.067901} and quantum simulation \cite{doi:10.1126/science.1163861}, to name a few. For many of the envisaged studies and applications, initial preparation of the molecular sample in the rovibronic ground state, i.e., the lowest energy level of the electronic ground state, is desired \cite{Carr2009}.

In ongoing experiments, ultracold molecules are first created in a weakly-bound level via Feshbach resonance \cite{RevModPhys.82.1225}. For collisional and radiative stability, Feshbach molecules are then quickly moved to their absolute ground state \cite{10.1063/1.4916903}. During this process, at least one intermediate excited state should be introduced as a bridge between the Feshbach and the ground state, which should have favorable transition dipole moments and good Frank-Condon overlaps with the vibrational wavefunctions of the Feshbach and deeply-bound ground states.
The common technique for ground state transfer is the well-known stimulated Raman adiabatic passage (STIRAP) \cite{RevModPhys.89.015006, doi:10.1126/science.aau7230, 10.1063/5.0046194, 10.1063/5.0082309}, and its variants, such as multi-state chainwise-STIRAP (C-STIRAP) \cite{Danzl2010, Mark2009, PhysRevA.78.021402, PhysRevA.56.4929, PhysRevA.44.7442}. The successes of conventional STIRAP (C-STIRAP) relies on the existence of the dark (adiabatic transfer) state, which should be followed adiabatically \cite{10.1063/1.471424, 10.1063/1.458514, 10.1063/1.4818526, PhysRevA.102.023515}. To date, these techniques have been successfully demonstrated for species such as
homonuclear Rb$_2$, Cs$_2$, Sr$_2$ and Li$_2$ \cite{PhysRevLett.98.043201, PhysRevLett.101.133005, Danzl1062, Leung_2021, Danzl_2009, PhysRevLett.109.115302, PhysRevA.102.013310},
as well as heteronuclear KRb, RbCs, NaK, RbSr, NaRb, NaLi, RbHg, LiK,
and NaCs \cite{PhysRevLett.105.203001, Ospelkaus2008, PhysRevLett.113.255301, PhysRevLett.113.205301, Ni231, PhysRevA.94.022507,
PhysRevA.97.013405, PhysRevLett.122.253201, PhysRevLett.114.205302,
PhysRevLett.116.205303, PhysRevLett.119.143001, PhysRevLett.124.133203, PhysRevLett.126.123402, PhysRevLett.130.113002, PhysRevResearch.4.L022019, Gregory_2015, Voges_2019, Warner_2023, PhysRevLett.125.083401, PhysRevA.97.013405, C1CP21769K, PhysRevA.96.063411}.

Since the transfer process is based on adiabatic passage, the process requires sufficiently high Rabi frequencies in order to suppress the non-adiabatic couplings and thereby to make adiabatic evolution possible \cite{RevModPhys.70.1003}. However, it may be problematic for molecular applications, this is due to very large laser intensities may breach various assumptions: other states will be coupled to the system, and multi-photon ionization or dissociation may be appreciable \cite{PhysRevLett.80.932, 10.1063/1.2180250}. In reality, due to the presence of losses induced by spontaneous and collide and the deviation from the dark (adiabatic transfer) state. The highest efficiencies for STIRAP and C-STIRAP transfer reported hitherto are around $90\%$ \cite{Christakis2023, PhysRevA.78.021402}. High transfer efficiency is essential in order to preserve the phase-space density of the ultracold mixture \cite{doi:10.1126/science.aau7230, Duda2023}. The imperfect efficiency results in a loss of phase space density, requiring further cooling of the molecules themselves.

To make the traditional adiabatic scheme more practical, a collection of methods referred as ``shortcut-to-adiabaticity" (STA) \cite{RevModPhys.91.045001, doi:10.1080/23746149.2021.1894978}, for instance, counter-diabatic driving (equivalently, transitionless quantum algorithm) \cite{Demirplak2003, Demirplak2005, doi:10.1063/1.2992152, Berry_2009, PhysRevLett.105.123003, PhysRevLett.109.100403, PhysRevLett.116.230503, Du2016, PhysRevA.94.063411, PhysRevA.89.033419, Li:17, Dou2021, 10.1063/5.0159448} and Invariant-based Inverse Engineering \cite{doi:10.1063/1.1664991, PhysRevA.86.033405, PhysRevA.83.062116, PhysRevA.99.013820, PhysRevA.96.023843} have been proposed. The main idea is to reach the same final state as the slow adiabatic evolution, without necessarily tracking the instantaneous adiabatic eigenstate \cite{RevModPhys.91.045001}.
In general, counter-diabatic driving needs to introduce extra coupling stems among ground states to cancel the non-adiabatic effect \cite{PhysRevLett.105.123003, PhysRevA.102.023515, Masuda2015}, however, which may not be a good solution for molecules due to different spin characters or weak Franck-Condon overlap between the molecular states that are involved \cite{Danzl2010}. Therefore, it is necessary to avoid the direct application of counter-diabatic pulses. Fortunately, in the frames of two- and three-level, ``Invariant-based Inverse Engineering'' has solved this issues by providing alternative STA, which is enticing because driving pulses have been designed with highly experimental feasibility. Its central idea is to utilize the Lewis-Riesenfeld (LR) invariants to carry the eigenstates of Hamiltonian from the initial to a desired final state, then to design the transient Hamiltonian via the LR invariant \cite{doi:10.1063/1.1664991}. This method maintains the same initial and final states as those in the adiabatic passage, but without following the adiabatic passage at the intermediate time instants.
Despite the advantages, the resulting efficiency of the invariant-based STA in standard $\Lambda$-type system may be extremely low due to the inevitable transient excitation of very short lived excited state \cite{PhysRevLett.125.193201, PhysRevA.96.013406, ZHANG2023106421, Zhang_2021, Ciamei_2017}. Therefore, the exploration of feasible Invariant-based STA methods for three- and multi-state systems, by reducing the number of applied pulses and by optimizing the pulse' shapes, deserves more investigations.

In this paper a theoretical method for the efficient and robust creation and detection of deeply-bound molecules based on the ``Invariant-based Inverse Engineering" is proposed. In present approach molecules are brought to the ground state through via a three-level $\Lambda$-type scheme and a five-level M-type scheme, respectively. Firstly, the major molecular losses are pre-suppressed by reducing the dynamics of three- and five-level molecular systems to those of effective two- and three-level counterparts under the large detuning condition. For effective two-level counterpart, I successively used two different protocols to modify the original fields.
The results show that both protocols give comparable performance and have good experimental feasibility, the transfer efficiency  can be improved drastically. For effective three-level counterpart, by setting a requirement towards the relation among the original incident pulses, this system will possess the simplest resonant couplings. Thereafter, this generalized model permits us to borrow the technique of ``Invariant-based Inverse Engineering''  recipe from three-state system into the five-level system for achieving robust and efficient population transfer. Finally, the required laser intensity and stability against parameter variation are also discussed. The proposed protocols offer a promising avenue for the preparation of gases with high-phase space density.

\section{\label{sec:level2}Model and Methods}
\subsection{Three-level $\Lambda$-type transfer scheme}

\begin{figure}[b]
\centering{\includegraphics[width=7cm]{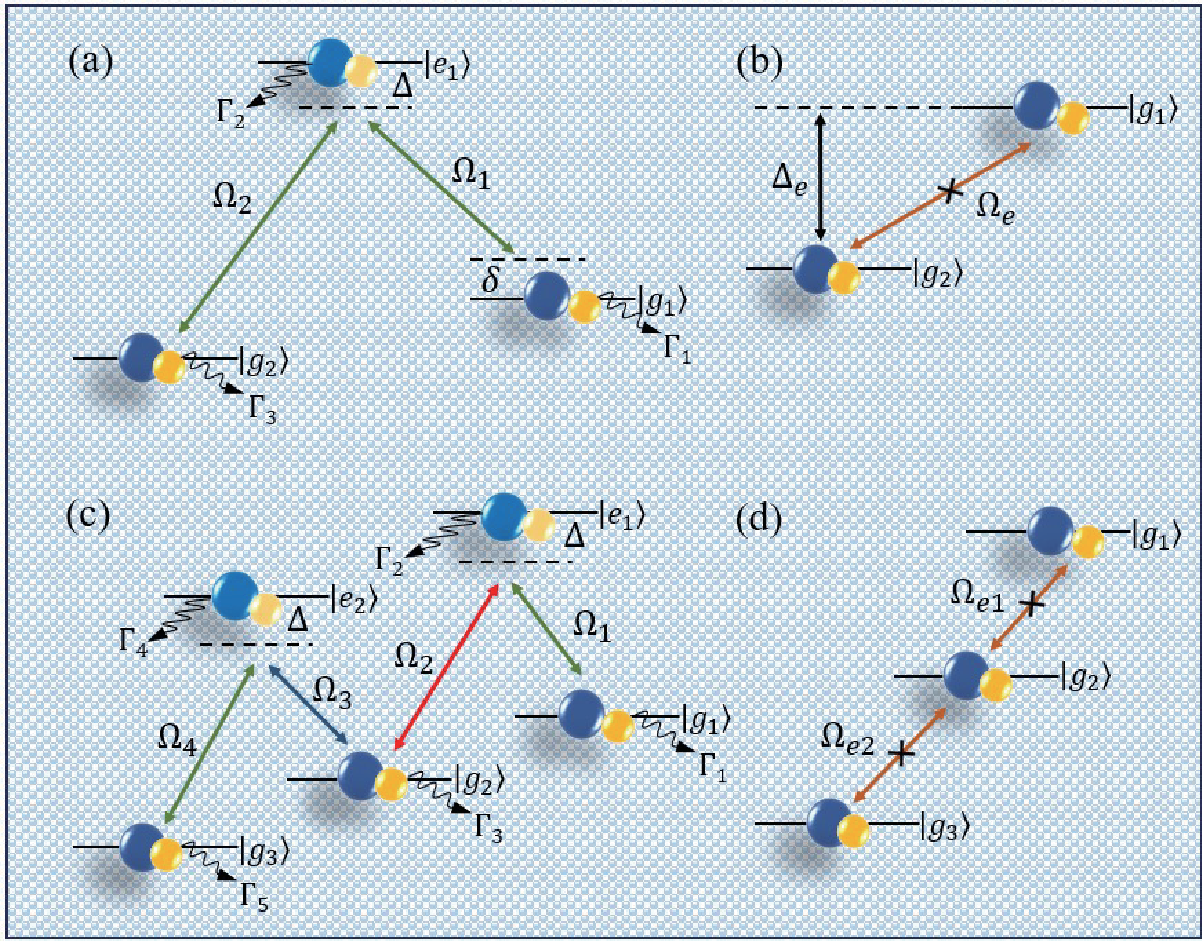}}
\caption{Schematic representations of coherent transfer ultracold molecule via (a) three-level $\Lambda$-type and (c) five-level M-type schemes. The effective (b) two- and (d) three-level counterparts after the adiabatic elimination. $\Gamma_i$ (i=1, 2, 3, 4, 5) represents the decay rate of corresponding level out of the system.}
\label{fig1}
\end{figure}
First consider a three-level ultracold molecular system in which both the initial Feshbach state $|g_1\rangle$ and final ground rovibrational state $|g_2\rangle$ are coupled by two lasers to a common intermediate excited state $|e_1\rangle$, as illustrated in Fig.~\ref{fig1}(a). The direct coupling between $|g_1\rangle$ and $|g_2\rangle$ is not considered infeasible, due to their different spin character and a negligible Franck-Condon factor, while excited level serves as a bridge is assumed to have favourable transition dipole moments and good Frank-Condon overlaps with both $|g_1\rangle$ and $|g_2\rangle$ vibrational wavefunctions.

The total molecular wave function can be expanded as
\begin{eqnarray}\label{1}
| \psi(t)\rangle=c_{1}(t)| g_1\rangle+c_{2}(t)| e_1\rangle+c_{3}(t)| g_2\rangle,
\end{eqnarray}
the vector $c_{1}(t), c_{2}(t)$ and $c_{3}(t)$ are the probability amplitudes of the corresponding state.
The evolution is then governed by the time-dependent Schr\"{o}dinger equation:
\begin{eqnarray}\label{2}
i\hbar\displaystyle\frac{\partial}{\partial t}c(t)=H(t)c(t).
\end{eqnarray}
In the interaction representation and after adopting the rotating-wave approximation, this system can be quantitatively described in terms of a three-level Hamiltonian $(\hbar=1)$
\begin{equation}\label{3}
H(t)=
\begin{bmatrix}
0&\Omega_{1}/2&0\\
\Omega_{1}/2&\Delta&\Omega_{2}/2\\
0&\Omega_{2}/2&\delta\\
\end{bmatrix}.
\end{equation}
In which $\Omega_{1}$ and $\Omega_{2}$ are Rabi frequencies of pump and Stokes laser fields. The quantities $\Delta$ and $\delta$ stand for single- and two-photon detunings, respectively.
Here two Rabi frequencies are assumed to be equal valued and simultaneous, i.e., $\Omega=\Omega_1=\Omega_2$. Then Hamiltonian can be reduced to
\begin{equation}\label{4}
H(t)=
\begin{bmatrix}
0&\Omega/2&0\\
\Omega/2&\Delta&\Omega/2\\
0&\Omega/2&\delta\\
\end{bmatrix}.
\end{equation}
In the regime of large single-photon detuning ($\Delta$ $\gg$ $\Omega, \delta$), level state $| e_1\rangle$ is scarcely
populated during the population transfer \cite{PhysRevX.9.021039}. After adiabatically eliminating (AE) the excited state $| e_1\rangle$, the effective Hamiltonian can be given by
\begin{equation} \label{5}
H_{e}(t)=
\begin{bmatrix}
\Delta_{e}/2&\Omega_{e}/2\\
\Omega_{e}/2&-\Delta_{e}/2\\
\end{bmatrix},
\end{equation}
in which
\begin{eqnarray}
\Omega_{e}&&=-\frac{\Omega^2}{2\Delta}, \label{6}\\ \nonumber\\
\Delta_{e}&&=-\delta,\label{7}
\end{eqnarray}
then one-photon transitions to the excited state $|e_1\rangle$ are negligible and two-photon transitions directly to the ground
molecular state $|g_2\rangle$ dominate, as illustrated in Fig.~\ref{fig1}(b).

If the evolution is adiabatic, a perfect population transfer can be achieved. Viewed mathematically, it requires the coupling between the adiabatic states can be negligible compared with the difference between their eigenfrequencies \cite{10.1063/1.481829}, explicitly,
\begin{equation}\label{8}
\frac{1}{2}|\dot\Omega_{e}{\Delta}_{e}-\dot\Delta_{e}{\Omega}_{e}| \ll (\Omega_{e}^2+\Delta_{e}^2)^{3/2}.
\end{equation}
However, the adiabatic requirements are not always affordable in reality. In order to release the adiabatic requirements and maintain the high transfer efficiency, one can apply the ``STA'' methods. For example, Hamiltonian $H_{e}(t)$ can be directly compatible with ``Invariant-based Inverse Engineering'' recipe, one can alternatively obtain a non-adiabatic population transfer.

Associated with the time-dependent Hamiltonian are Hermitian dynamical invariants $I(t)$, satisfying that $
dI(t)/dt\equiv\partial I(t)/\partial t +(1/ i\hbar)[I(t),H_{e}(t)]=0$,
so that their expectation values remain constant. The $I(t)$ can parameterize to  \cite{PhysRevA.86.033405}
\begin{equation}\label{9}
I(t)=\frac{\Omega_0}{2}
\begin{bmatrix}
\cos\theta&\sin\theta e^{-i\beta}\\
\sin\theta e^{i\beta}&-\cos\theta\\
\end{bmatrix}.
\end{equation}
where $\Omega_0$ is an arbitrary constant with units of frequency to keep $I(t)$ with dimensions of energy, and the time-dependent auxiliary parameters $\theta$ and $\beta$ satisfy the relation
\begin{eqnarray}
\dot{\theta}&&=-\tilde{\Omega}_{e}\sin\beta,\label{10}\\ \nonumber\\
\dot{\beta}&&=-\tilde{\Omega}_{e}\cot\theta\cos\beta-\tilde{\Delta}_{e}.\label{11}
\end{eqnarray}
The eigenstates of $I(t)$ have the same form as that of $H_{e}(t)$, which can be expressed as
\begin{eqnarray}
|\phi_{+}(t)\rangle&&=\cos(\frac{\theta}{2})e^{-i\beta}|g_1\rangle+\sin(\frac{\theta}{2})|g_2\rangle,\label{12}\\ \nonumber\\
|\phi_{-}(t)\rangle&&=\sin(\frac{\theta}{2})|g_1\rangle-\cos(\frac{\theta}{2})e^{i\beta}|g_2\rangle.\label{13}
\end{eqnarray}
Based on the Lewis-Riesenfeld theory, the solution of the time-dependent Schr\"{o}dinger equation, up to a global phase factor $\zeta$, can be expressed as $\psi(t)=\sum_na_ne^{i\zeta_n(t)}|\phi_{n}(t)\rangle$, where $\zeta_n(t)\equiv\int_{t_i}^{t}\langle\phi_{n}(t^{'})|i\frac{\partial}{\partial t^{'}}-H_{e}(t^{'})|\phi_{n}(t^{'})\rangle dt^{'}$, in our case $t_i=0$. To realize the transfer along eigenstate $|\phi_{+}(t)\rangle$, next I am ready to apply inverse engineering by means of two different protocols.

\subsubsection{Protocol 1}
In the first protocol, the parameter $\theta(t)$ can be set as $\theta(t)=\pi t/t_f$ as a particular case in order to satisfy $\theta(0)=0, \theta(t_f)=\pi$. According to the analysis of previous work, the optimum value of $\beta$ should be $\pi/2$ \cite{PhysRevA.88.033406}. However, this results in $\Delta_{e}=0$. To avoid that one can choose a value near $\pi/2$, for example, let $\beta=\pi/1.99$.

According to Eqs.~(\ref{10}) and~(\ref{11}) one can get the following relation
\begin{eqnarray}
\tilde{\Omega}_{e}&&=-\frac{\pi}{t_f\sin\beta},\label{14}\\ \nonumber\\
\tilde{\Delta}_{e}&&=-\tilde{\Omega}_{e}\cot\theta\cos\beta.\label{15}
\end{eqnarray}
To eliminate $|g_1\rangle$$\leftrightarrow$$|g_3\rangle$ coupling, one can further replace $\Omega(t)$ and $\delta$ in Hamiltonian (\ref{4}) with modified ones. Like Eqs.~(\ref{6}) and~(\ref{7}), one can impose
\begin{eqnarray}
\tilde{\Omega}_{e}&&=-\frac{\tilde{\Omega}^2}{2\Delta},\label{16}\\ \nonumber\\
\tilde{\Delta}_{e}&&=-\tilde{\delta}.\label{17}
\end{eqnarray}
After going back to the basis $\{|g_1\rangle, |e_1\rangle, |g_2\rangle\}$, the modified fields and modified two-photon detuning can be calculated inversely as
\begin{eqnarray}
\tilde{\Omega}&&=\sqrt{\frac{2\Delta\pi}{t_f\sin\beta}},\label{18}\\ \nonumber\\
\tilde{\delta}&&=-\frac{\pi}{t_f}\frac{\cot\theta}{\cot\beta}.\label{19}
\end{eqnarray}
For protocol 1, the modified Hamiltonian can be written as
\begin{equation}\label{20}
\tilde{H}(t)=
\begin{bmatrix}
0&\tilde{\Omega}/2&0\\
\tilde{\Omega}/2&\Delta&\tilde{\Omega}/2\\
0&\tilde{\Omega}/2&\tilde{\delta}\\
\end{bmatrix}.
\end{equation}
\subsubsection{Protocol 2}
In the second protocol, $\theta$ can be set as $\theta(0)=0, \theta(t_f)=\pi$ and $\dot{\theta}(0)=\dot{\theta}(t_f)=0$. The former is used to guarantee the desired initial and final states; the latter makes $\tilde{\Omega}_{e}(0)=\tilde{\Omega}_{e}(t_f)=0$. According to the ansatz $\theta(t)=\sum_{k=0}^{3}a_kt^k$, thus one can obtain $\theta(t)=\frac{3\pi}{{t_f}^2}t^2-\frac{2\pi}{{t_f}^3}t^3.$
For $\beta=\pi/2$ exactly, then $\Delta_{e}=0$. Similarly, according to Eqs.~(\ref{16}) and~(\ref{17}), the modified fields can be calculated inversely as
\begin{eqnarray}
\tilde{\Omega}&&=\sqrt{2\Delta\dot{\theta}}=\sqrt{\Delta(\frac{12\pi}{{t_f}^2}t-\frac{12\pi}{{t_f}^3}t^2)},\label{21}\\ \nonumber\\
\tilde{\delta}&&=0.\label{22}
\end{eqnarray}
For protocol 2, the modified Hamiltonian can be written as
\begin{equation}\label{23}
\tilde{H}(t)=
\begin{bmatrix}
0&\tilde{\Omega}/2&0\\
\tilde{\Omega}/2&\Delta&\tilde{\Omega}/2\\
0&\tilde{\Omega}/2&0\\
\end{bmatrix}.
\end{equation}
With the well-designed $\tilde{\Omega}$ and $\tilde{\delta}$, the system in principle inherits many of the interesting features of ``IIE'', but due to the practical limitations, e.g., the laser intensity, the performance would depend on those limitations.

In order to verify the validity of our protocols, we are going to study the evolution of the $^{87}$Rb$_2$ system by numerical calculations \cite{PhysRevLett.98.043201}. I also investigate more restrictive conditions such as low intensity lasers.  
Since the molecules face the decays due to spontaneous emission and collisions \cite{PhysRevA.85.023629, PhysRevA.87.043631}, the decays from states displayed in Fig.~\ref{fig1}(a) should be taken into account. Thus, the evolution of the system is governed by the Liouville-von Neumann equation:
\begin{eqnarray}\label{24}
i\hbar\displaystyle\frac{d\rho}{dt}=[\tilde{H}, \rho]+\Gamma\rho,
\end{eqnarray}
in which $\rho$ is the density operator and $\Gamma$ represents phenomenological decay rates.

\begin{figure}[t]
\centering{\includegraphics[width=8.5cm]{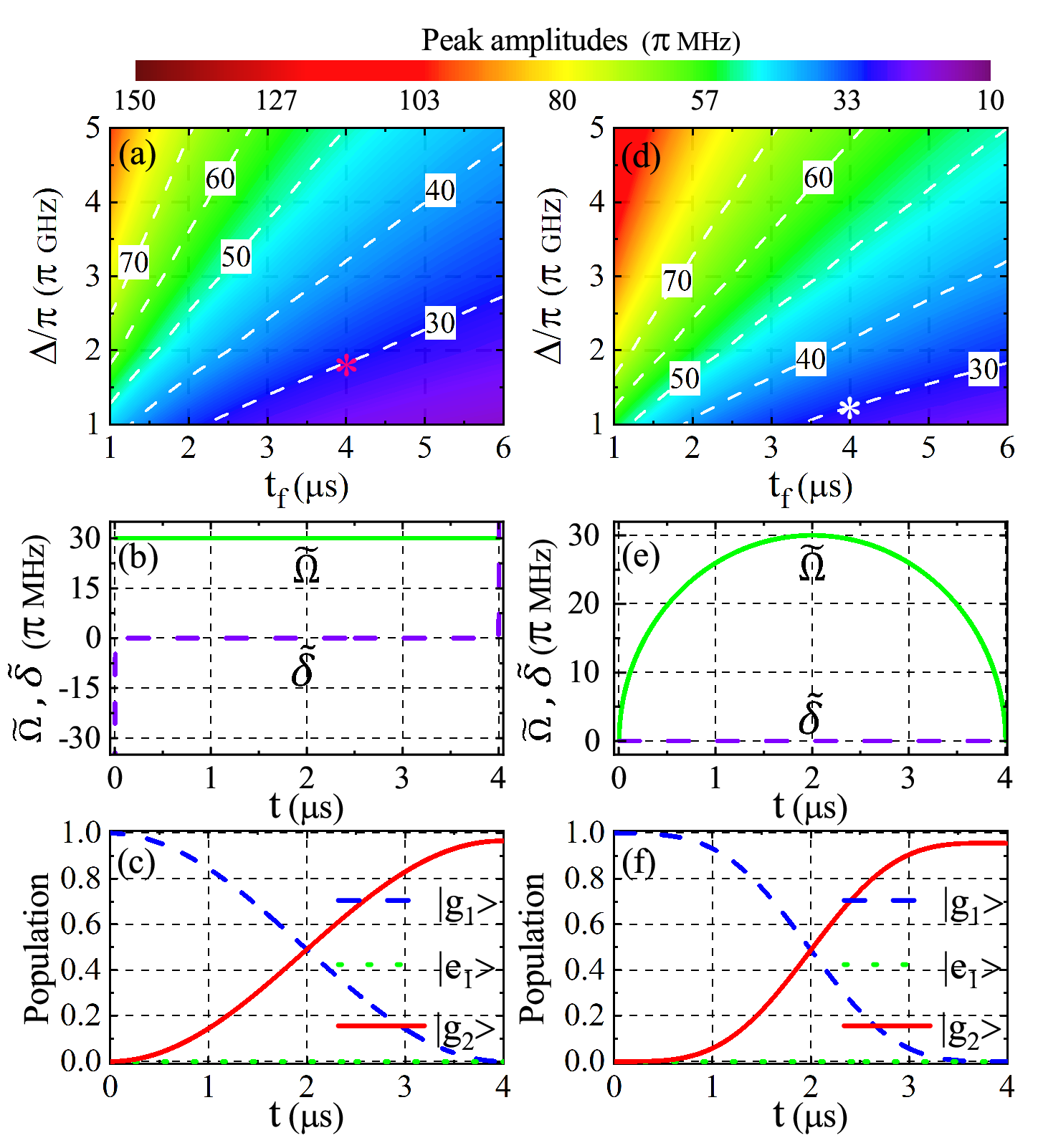}}
\caption{(a) and (b) represent the maximum amplitudes of $\tilde{\Omega}$ as functions of $t_f$ and $\Delta$ under protocol 1 and protocol 1, respectively. The pink star points to the parameter value used in (b) and (c), i.e., $t_f=4\mu s, \Delta=1.8\pi{\rm{GHz}}$; The white star points to the parameter value used in (e) and (f), i.e., $t_f=4\mu s, \Delta=1.2\pi{\rm{GHz}}$, which illustrate the modified pulses and the efficiencies under protocol 1 and protocol 1, respectively. Parameters used: $\Gamma_1=2\pi\times0.72{\rm{KHz}}$, $\Gamma_2=2\pi\times12{\rm{MHz}}$, $\Gamma_3=2\pi\times0.4{\rm{KHz}}$.}
\label{fig2}
\end{figure}
The upper frames of Fig.~\ref{fig2} show the maximum value of laser intensity $\tilde{\Omega}_{max}$ of (a) protocol 1 and (d) protocol 2 as function of operation time $t_f$ and detuning $\Delta$, respectively.
The results reveal that longer operation time is required to reduce the laser intensity when $\Delta$ is fixed, it is reasonable because the relation between energy (laser intensity) and operation time, in general, satisfies $\Omega_{max}$ $\varpropto$ $1/t_f$. That is to say, although in principle one can reproduce the dynamics in infinitely short time, here the limitations on laser intensity will restrict the process up to a time limit. Besides, we point out that the laser intensity can be further decreased by using a smaller detuning $\Delta$. This is because the modified $\tilde{\Omega}$ is proportional to the detuning, as can be found in Eqs.~(\ref{18}) and~(\ref{21}).

When an allowed maximum value of laser intensity $\tilde{\Omega}_{max}$ is fixed, for example, if $\tilde{\Omega}_{max}=30\pi{\rm{MHz}}$, one can assume the detuning required for protocol 1 and protocol 2 are $\Delta=1.8\pi{\rm{GHz}}$ and $\Delta=1.2\pi{\rm{GHz}}$ to meet this condition, respectively. Note that the time required for both protocols is $t_f = 4\mu s$. Fig.~\ref{fig2}(b) shows the modified coupling $\tilde{\Omega}$ and the modified two-photon detuning $\tilde{\delta}$ under the protocol 1. Note that the the modified coupling $\tilde{\Omega}$ is independent of time. The corresponding population transfer is displayed in
Fig.~\ref{fig2}(c). Fig.~\ref{fig2}(e) shows the modified coupling $\tilde{\Omega}$ under the protocol 2, which have the time profiles that are more accessible in experiment. The corresponding population transfer is displayed in Fig.~\ref{fig2}(f).
As expected, the results show indeed that both protocols drastically enhances the transfer efficiencies. During the transfer process, AE protocol ensures the decoupling of the excited states from the dynamics, we can directly move the population from state $|g_1\rangle$ to $|g_2\rangle$, $|e_1\rangle$ is only used to induce transitions but never significantly populated. Thus the transfer process is insensitive to the properties of excited molecules, e.g., fast spontaneous emission.
\begin{figure}[t]
\centering{\includegraphics[width=6cm]{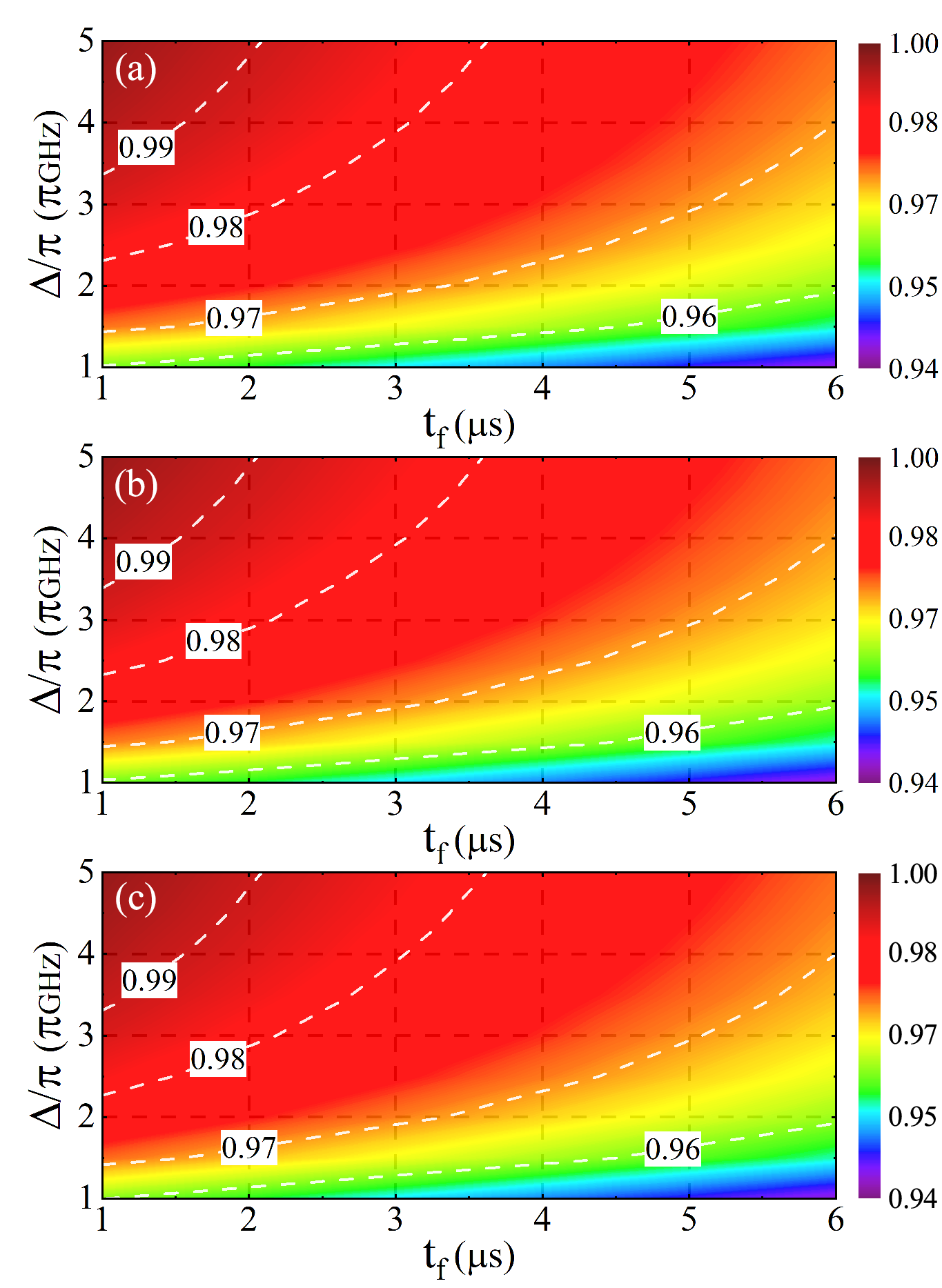}}
\caption{The transfer efficiencies for (a) protocol 1 with modified two-photon detuning, (b) protocol 1 without modified two-photon detuning and (c) protocol 2 as functions of $t_f$ and $\Delta$. Parameters used: $t_f\in[1, 6]\mu s$, $\Delta\in[1, 5]\pi{\rm{GHz}}$, all the other parameters are the same as those in Fig.~\ref{fig2}.}
\label{fig3}
\end{figure}

It should be noted that the modified two-photon detuning $\tilde{\delta}$ under the protocol 1 have singularities, this renders that physical realization of the Hamiltonian~(\ref{20}) is not easy. In fact, the $\tilde{\delta}$ contributes very little to the dynamics. Further calculations in Fig.~\ref{fig3} can illustrate this abundantly, which shows a contour plot of the transfer efficiencies against the operation time $t_f$ the single-photon detuning $\Delta$. Fig.~\ref{fig3}(a) represents the exact implementation Hamiltonian~(\ref{20}). As a contrast, in Fig.~\ref{fig3}(b), the approximation of neglecting $\tilde{\delta}$ provides essentially the same dynamics as in Fig.~\ref{fig3}(a). Such an interesting phenomenon indicates that the protocol 1 can be simplified technically, requiring only two identical constant couplings. In order for a complete discussion, Fig.~\ref{fig3}(c) shows the exact implementation Hamiltonian~(\ref{23}). As it is observed, for the current cases, all the protocols give comparable performances and both protocols have high experimental feasibility.

\begin{figure}[t]
\centering{\includegraphics[width=10cm]{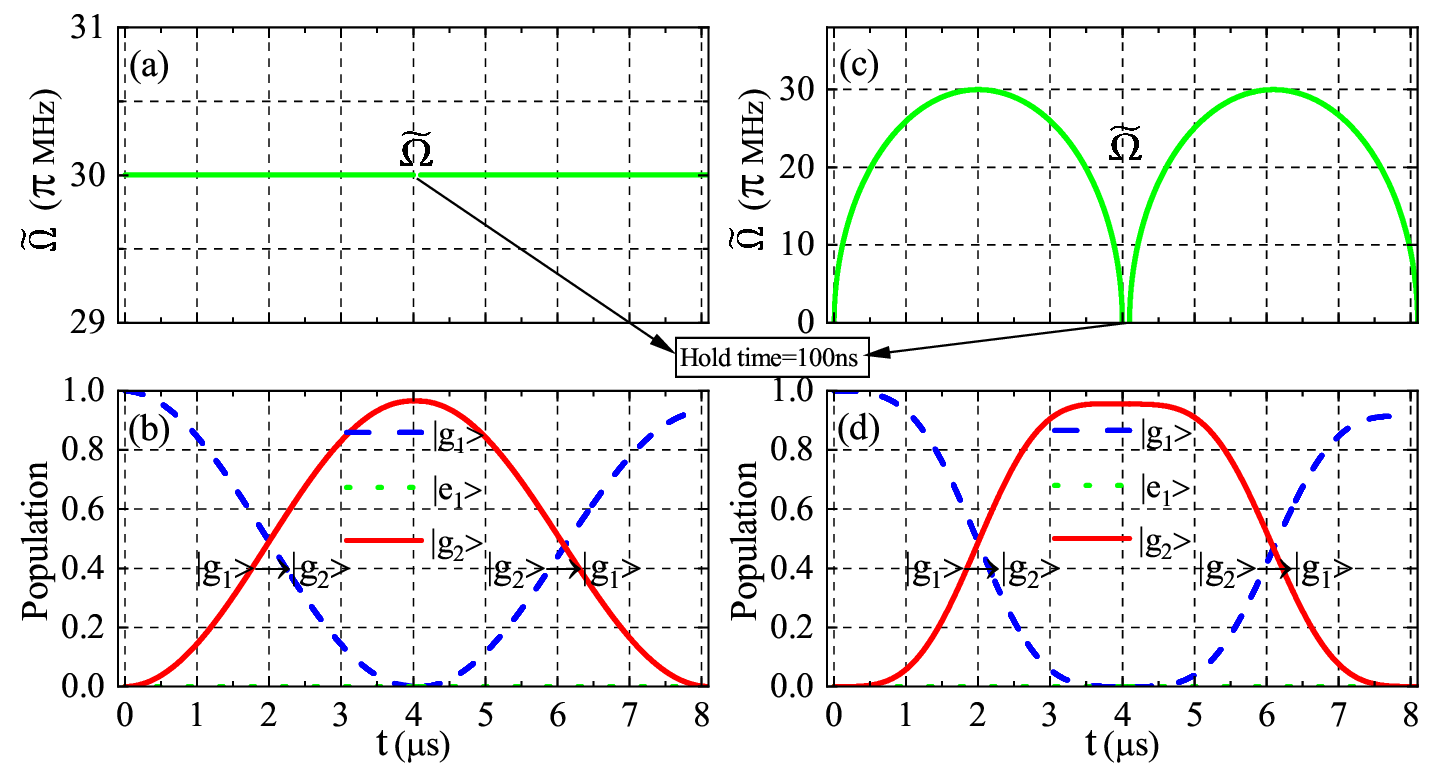}}
\caption{(a) and (c) illustrate the corresponding modified $\tilde{\Omega}$ of protocol 1 and the protocol 2, which are used for the creation and detection of the ultracold deeply-bound molecules. (b) and (d) show the round-trip transfer efficiency of protocol 1 and protocol 2, respectively. All parameters are the same as those in Fig.~\ref{fig2}(c) and (f).}
 \label{fig4}
\end{figure}

To determine the efficiency of our protocols, the molecules should be transferred back to the Feshbach state and detected by
standard absorption imaging technique. The efficiency of back-transfer directly impacts the detection fidelity. High detection
fidelity will be critical for the preparation and characterization of novel many-body phases of dipolar molecules, such as crystalline bulk phases \cite{PhysRevLett.98.060404}, exotic density order \cite{Baranov2012}, or spin order in optical lattices \cite{PhysRevLett.107.115301}. Generally, the back-transfer is realized by reversing the time sequence of the incident pulses for the creation process. Instead, the present protocols do not require a change in the laser sequence, this is due to AE results in a symmetry of the eigenstates of the invariant $I(t)$, which only involves an unimportant change of sign in exponential term. To better illustrate this point, Fig.~\ref{fig4} directly gives corresponding examples of the detections following the creation results in Fig.~\ref{fig2}(c) and (f). The left and right columns of Fig.~\ref{fig4} show the detection results of protocol 1 and protocol 2, respectively. Obviously, the first step one needs to do is to perform a one-way transfer, which would nominally convert Feshbach molecules to the rovibrational ground state. After a holding time $100ns$, the round-trip transfer is switched on, which would nominally bring the population back. In these figures, one can clearly see the round-trip transfer of population with perfect efficiency, which implies high efficiencies of the detection. The reason can be attributed to the fact that the one-way transfer follows exactly the path laid down by the $|\phi_{+}(t)\rangle$ state, while the round-trip transfer follows exactly the path laid down by the $|\phi_{-}(t)\rangle$ state.

\subsection{Five-level M-type transfer scheme}
If the typical STIRAP in $\Lambda$-type system does not work due to the weak Franck-Condon overlap between the molecular states that are involved, I also considers a M-type chainwise stimulated Raman adiabatic passage (C-STIRAP) transfer scheme, as illustrated in Fig.~\ref{fig1}(c), in which both the initial Feshbach state $|g_1\rangle$ and final ground rovibrational state $|g_3\rangle$ are coupled through a series of intermediate vibrational states. Here, two excited level $|e_1\rangle$ and $|e_2\rangle$ serve as the bridges are required, one having a good Franck-Condon overlap with $|g_3\rangle$, and the other with the initial Feshbach molecular state $|g_1\rangle$. In this case the initial and final vibrational levels of each step do not differ significantly.

The total molecular wave function can be expanded as
\begin{eqnarray}\label{25}
| \psi(t)\rangle=c_{1}(t)| g_{1}\rangle&&+c_{2}(t)| e_{1}\rangle+c_{3}(t)| g_{2}\rangle\nonumber \\
&&+c_{4}(t)| e_{2}\rangle+c_{5}(t)| g_{3}\rangle.
\end{eqnarray}
Under the rotating wave approximation, this system can be quantitatively described in terms of a five-level Hamiltonian $(\hbar=1)$,
\begin{equation}\label{26}
H(t)=
\begin{bmatrix}
0&\Omega_{1}/2&0&0&0\\
\Omega_{1}/2&\Delta&\Omega_{2}/2&0&0\\
0&\Omega_{2}/2&0&\Omega_{3}/2&0\\
0&0&\Omega_{3}/2&\Delta&\Omega_{4}/2\\
0&0&0&\Omega_{4}/2&0\\
\end{bmatrix}.
\end{equation}
Similarly, one can also exclude direct population of the excited states by AE protocol, then two-photon transitions $|g_1\rangle$$\leftrightarrow$$|g_2\rangle$ and $|g_2\rangle$$\leftrightarrow$$|g_3\rangle$ will dominate, as illustrated in Fig.~\ref{fig1}(d). This leads to an equation characterizing an effective off-resonant three-state system given by
\begin{equation}\label{27}
H_e(t)=
\begin{bmatrix}
\Delta_{e_1}&\Omega_{e_1}&0\\
\Omega_{e_1}&\Delta_{e_2}&\Omega_{e_2}\\
0&\Omega_{e_2}&\Delta_{e_3}\\
\end{bmatrix},
\end{equation}
where the effective couplings are defined as
$\Omega_{e_1}=-\Omega_{1}\Omega_{2}/(4\Delta), \Omega_{e_2}=-\Omega_{3}\Omega_{4}/(4\Delta)$, respectively.
Three diagonal elements are
$\Delta_{e_1}=-\Omega_{1}^2/(4\Delta),
\Delta_{e_2}=-(\Omega_{2}^2+\Omega_{3}^2)/(4\Delta),
\Delta_{e_3}=-\Omega_{4}^2/(4\Delta)$. Obviously, the system after AE subjects to dynamic Stark shifts from the trapping light, which can be expected to reduce the transfer efficiency. In order to make STIRAP in the strict sense possible, one first need to assume that the three diagonal elements are equal to each other,
i.e., $\Delta_{e_1}=\Delta_{e_2}=\Delta_{e_3}=\Delta$.
This requires that the four Rabi frequencies $\Omega_{j} (j=1,2,3,4)$ should satisfy
\begin{eqnarray}\label{28}
\Omega_{1}=\Omega_{4}=\sqrt{\Omega_{2}^2+\Omega_{3}^2}.
\end{eqnarray}
Note that this kind of pulse sequence is different from previous the straddling STIRAP \cite{PhysRevA.56.4929} and the alternating STIRAP scheme \cite{PhysRevA.44.7442} (which are two possible versions of STIRAP for multilevel systems with odd number of levels).
By further setting $c_j =c^{'}_{j}e^{-i\Delta t}(i=1, 3, 5)$, the above equation can be written in the following
\begin{eqnarray}\label{29}
i\frac{d}{dt}
\begin{bmatrix}
c^{'}_1\\
c^{'}_3\\
c^{'}_5\\
\end{bmatrix}
=
\begin{bmatrix}
0&\Omega_{e_1}&0\\
\Omega_{e_1}&0&\Omega_{e_2}\\
0&\Omega_{e_2}&0\\
\end{bmatrix}
\begin{bmatrix}
c^{'}_1\\
c^{'}_3\\
c^{'}_5\\
\end{bmatrix}.
\end{eqnarray}
Thus one can obtain that
\begin{equation}\label{30}
H_e^{'}(t)
\begin{bmatrix}
0&\Omega^{'}_{e_1}/2&0\\
\Omega^{'}_{e_1}/2&0&\Omega^{'}_{e_2}/2\\
0&\Omega^{'}_{e_2}/2&0\\
\end{bmatrix}.
\end{equation}
where
\begin{eqnarray}
\Omega^{'}_{e_1}&&=-\frac{\Omega_{2}\sqrt{\Omega_{2}^2+\Omega_{3}^2}}{2\Delta},\label{31}\\ \nonumber\\ \Omega^{'}_{e_2}&&=-\frac{\Omega_{3}\sqrt{\Omega_{2}^2+\Omega_{3}^2}}{2\Delta}.\label{32}
\end{eqnarray}
Hence the five-level system can be generalized into an effective $\Lambda$-type structure with the simplest resonant coupling, as illustrated in Fig.~\ref{fig1}(d). It is similar to STIRAP and allows for complete STIRAP-like population transfer \cite{zhang2023highly}.

Similarly, based on the invariant-based STA, the dynamical invariant $I(t)$ of
the Hamiltonian $H_e^{'}(t)$
, fulfilling the relation $
dI/dt\equiv\partial I/\partial t +(1/ i\hbar)[I(t),H_e^{'}(t)]=0$, can be expressed as
\begin{equation}\label{33}
I=\frac{\Omega_0}{2}
\begin{bmatrix}
0&\cos\chi\sin\vartheta&-i\sin\chi\\
\cos\chi\sin\vartheta&0&\cos\chi\cos\vartheta\\
i\sin\chi&\cos\chi\cos\vartheta&0\\
\end{bmatrix}.
\end{equation}
The eigenstates of I(t) can be expressed as
\begin{eqnarray}\label{34}
|\phi'_{0}(t)\rangle&&=
\begin{bmatrix}
\cos\chi\cos\vartheta\\
-i\sin\chi\\
-\cos\chi\sin\vartheta\\
\end{bmatrix},
\end{eqnarray}
and
\begin{eqnarray}\label{35}
|\phi'_{\pm}(t)\rangle&&=\frac{1}{\sqrt{2}}
\begin{bmatrix}
\sin\chi\cos\vartheta\pm i\sin\vartheta\\
i\cos\chi\\
-\sin\chi\sin\vartheta\pm i\cos\vartheta\\
\end{bmatrix}.
\end{eqnarray}
The time-dependent parameters $\chi$ and $\vartheta$ are related to Rabi frequencies $\Omega^{'}_{e_{1,2}}$ as follows,
\begin{eqnarray}
\Omega^{'}_{e_{1}}&&\equiv\tilde{\Omega}_{e_{1}}=2(\dot{\vartheta}\cot\chi\sin\vartheta+\dot{\chi}\cos\vartheta),\label{36}\\ \nonumber\\
\Omega^{'}_{e_{2}}&&\equiv\tilde{\Omega}_{e_{2}}=2(\dot{\vartheta}\cot\chi\cos\vartheta-\dot{\chi}\sin\vartheta).\label{37}
\end{eqnarray}
The boundary conditions of $\chi$ and $\vartheta$ are set as $\chi(0)=\varepsilon, \chi(t_f)=\varepsilon, \chi(t_f/2)=\pi/4,$ and $\vartheta(0)=0, \vartheta(t_f)=\pi/2,$ here a small value $\varepsilon$ has been introduced to avoid the infinite Rabi frequencies at the initial time $t=0$ and final time $t=t_f$~\cite{PhysRevA.86.033405}. Meanwhile, to keep the state at initial and final times stationary, $\chi$ and $\vartheta$ also need to satisfy $\dot{\chi}(0)=0, \dot{\chi}(t_f)=0$ and $\dot{\vartheta}(0)=0, \dot{\vartheta}(t_f)=0.$ Here $\chi(t)$ and $\vartheta(t)$ can be designed as polynomial ansatz: $\chi(t)=\sum_{j=0}^4a_jt^j$, $\vartheta(t)=\sum_{j=0}^3b_jt^j$.

Then the form of the modified Hamiltonian is
\begin{equation}\label{38}
\tilde{H}_e(t)=
\begin{bmatrix}
0&\tilde{\Omega}_{e_{1}}/2&0\\
\tilde{\Omega}_{e_{1}}/2&0&\tilde{\Omega}_{e_{1}}/2\\
0&\tilde{\Omega}_{e_{1}}/2&0
\end{bmatrix}.
\end{equation}
Now let us go back to the five-level system and design the modified fields. Like Eqs.~(\ref{31}) and~(\ref{32}), one can impose
\begin{eqnarray}
\tilde{\Omega}_{e_{1}}&&=-\frac{\tilde{\Omega}_{2}\sqrt{\tilde{\Omega}_{2}^2+\tilde{\Omega}_{3}^2}}{2\Delta}, \label{39}\\ \nonumber\\ \tilde{\Omega}_{e_{2}}&&=-\frac{\tilde{\Omega}_{3}\sqrt{\tilde{\Omega}_{2}^2+\tilde{\Omega}_{3}^2}}{2\Delta}.\label{40}
\end{eqnarray}
By inversely deriving, the explicit expressions of $\tilde{\Omega}_{2}$ and $\tilde{\Omega}_{3}$ are obtained as follows:
\begin{eqnarray}
\tilde{\Omega}_{2}&&=\tilde{\Omega}_{e_{1}}\left(\frac{4\Delta^2}{\tilde{\Omega}^2_{e_{1}}+\tilde{\Omega}^2_{e_{2}}}\right)^{\frac{\rm{1}}{\rm{4}}},\label{41}\\ \nonumber\\
\tilde{\Omega}_{3}&&=\tilde{\Omega}_{e_{2}}\left(\frac{4\Delta^2}{\tilde{\Omega}^2_{e_{1}}+\tilde{\Omega}^2_{e_{2}}}\right)^\frac{1}{4}.\label{42}
\end{eqnarray}
According to Eq.~(\ref{28}), one can impose
\begin{eqnarray}\label{43}
\tilde{\Omega}_{1, 4}=\sqrt{\tilde{\Omega}_{2}^2+\tilde{\Omega}_{3}^2},
\end{eqnarray}
and calculate inversely the modified fields as
\begin{eqnarray}\label{44}
\tilde{\Omega}_{1, 4}=\left[4\Delta^2(\tilde{\Omega}^2_{e_{1}}(t)+\tilde{\Omega}^2_{e_{2}}(t))\right]^{\frac{1}{4}}.
\end{eqnarray}
With modified pulses $\tilde{\Omega}_{1-4}$, the five-level molecular Hamiltonian can be rewritten as
\begin{equation}\label{45}
\tilde{H}(t)=
\begin{bmatrix}
0&\tilde{\Omega}_{1}/2&0&0&0\\
\tilde{\Omega}_{1}/2&\Delta&\tilde{\Omega}_{2}/2&0&0\\
0&\tilde{\Omega}_{2}/2&0&\tilde{\Omega}_{3}/2&0\\
0&0&\tilde{\Omega}_{3}/2&\Delta&\tilde{\Omega}_{4}/2\\
0&0&0&\tilde{\Omega}_{4}/2&0\\
\end{bmatrix}.
\end{equation}
With the four modified Rabi frequencies, one can continue to carry out numerical calculations together with Eq.~(\ref{24}) to verify the feasibility of our method. We still take the parameters of molecule $^{87}$Rb$_{2}$ in the calculations \cite{PhysRevA.78.021402}.

\begin{figure}[t]
\centering{\includegraphics[width=7.5cm]{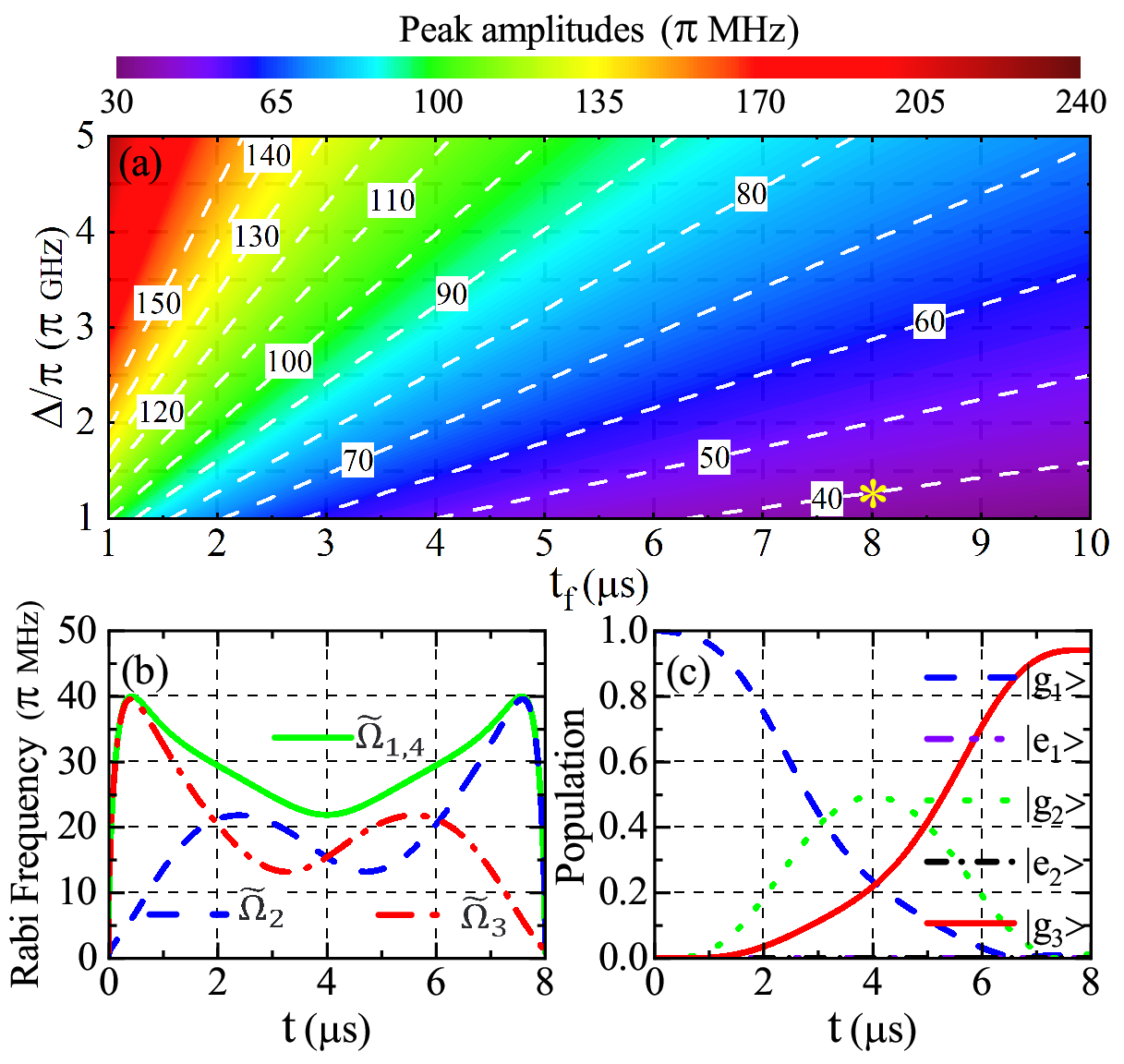}}
\caption{(a) The maximum amplitudes of $\tilde{\Omega}_{1, 4}$ as functions of $t_f$ and $\Delta$. The yellow star points to the parameter value used in (b) and (c), which illustrate the modified pulses and the efficiencies, respectively. Parameters used: $t_f =8\mu s$, $\Delta=1.27\pi{\rm{GHz}}$, $\varepsilon=0.03$, $\Gamma_1=\Gamma_3=0.01{\rm{MHz}}$, $\Gamma_2=\Gamma_4=30{\rm{MHz}}$, $\Gamma_5=0$.}
\label{fig5}
\end{figure}

\begin{figure}[t]
\centering{\includegraphics[width=7.5cm]{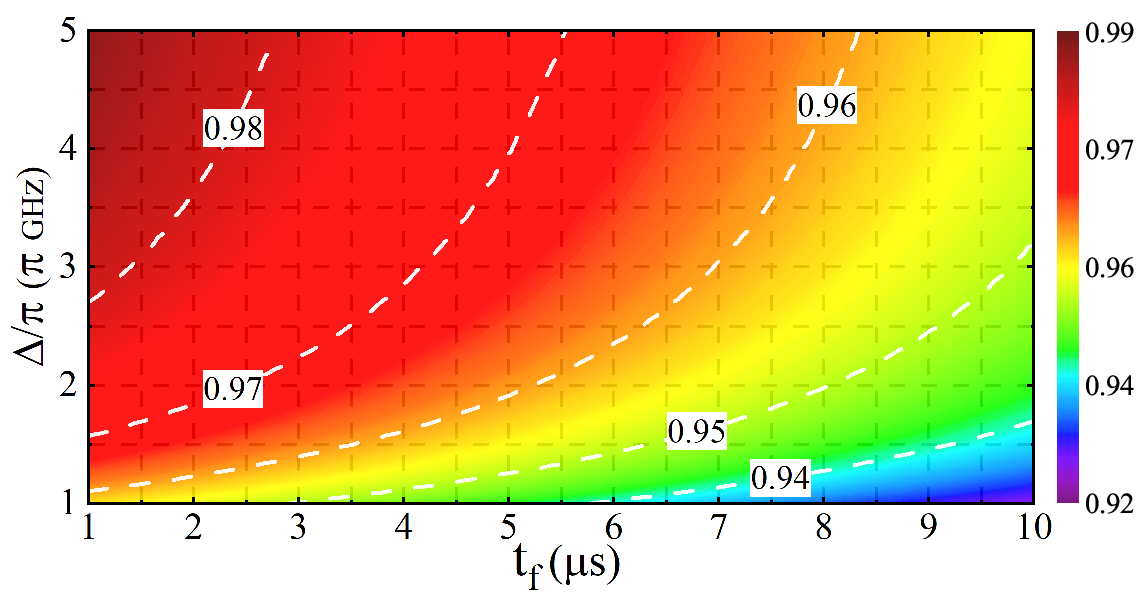}}
\caption{Transfer efficiency as functions of $t_f$ and $\Delta$. All the other parameters are the same as those in Fig.~\ref{fig5}.}
\label{fig6}
\end{figure}

Since the amplitudes of $\tilde{\Omega}_{2}$ and $\tilde{\Omega}_{3}$ are always smaller than those of $\tilde{\Omega}_{1, 4}$ (see Eq.~(\ref{43})). Therefore, it is sufficient to investigate the amplitudes of $\tilde{\Omega}_{1, 4}$ as function of operation time $t_f$ and detuning $\Delta$, as illustrated in Fig.~\ref{fig5}(a). The calculation show that the ways of reducing the laser intensity is the same as the previous three-level case. Clearly, the demand for maximum value of laser intensity decreases when the value of $t_f$ increases. For instance, one can choose a point in this figure when $\Delta=1.27\pi{\rm{GHz}}$ and $t_f=8\mu s$. They clearly show the feasibility of the present protocol that high transfer efficiency can be obtained with the maximum value of laser intensity are just around $40\pi{\rm{MHz}}$, as illustrated in Fig.~\ref{fig5}(b) and (c), which show corresponding time sequence of the four modified Rabi frequencies and the population evolution, respectively. Consequently, the initially very weakly-bound molecules are to be transferred in a few successive steps to the rovibrational ground state, acquiring more and more binding energy. Although the intermediate ground state $|g_2\rangle$ receives some transient populations along the way, it does not greatly influence on the final efficiency due to the long lifetime of this state.
Meanwhile, due to AE protocol, excited states $|e_1\rangle$ and $|e_2\rangle$ are only used to induce transitions but never significantly populated, the decay rates of them nearly do not influence the transfer efficiency. This will be very useful for depressing the effects of dissipation on the desired evolution of the system without relying on the adiabatic transfer state underlying the C-STIRAP.
Fig.~\ref{fig6} shows the transfer efficiency as functions of $t_f$ and $\Delta$, the result shows that even in presence of decay from these states, nearly perfect transfer efficiency can be obtained in this M-type system within a wide range of the
chosen parameters $t_f$ and $\Delta$. Therefore, these results validate the potential of our approach for preserving the phase-space density of the ultracold mixture.

\begin{figure}[t]
\centering{\includegraphics[width=7.5cm]{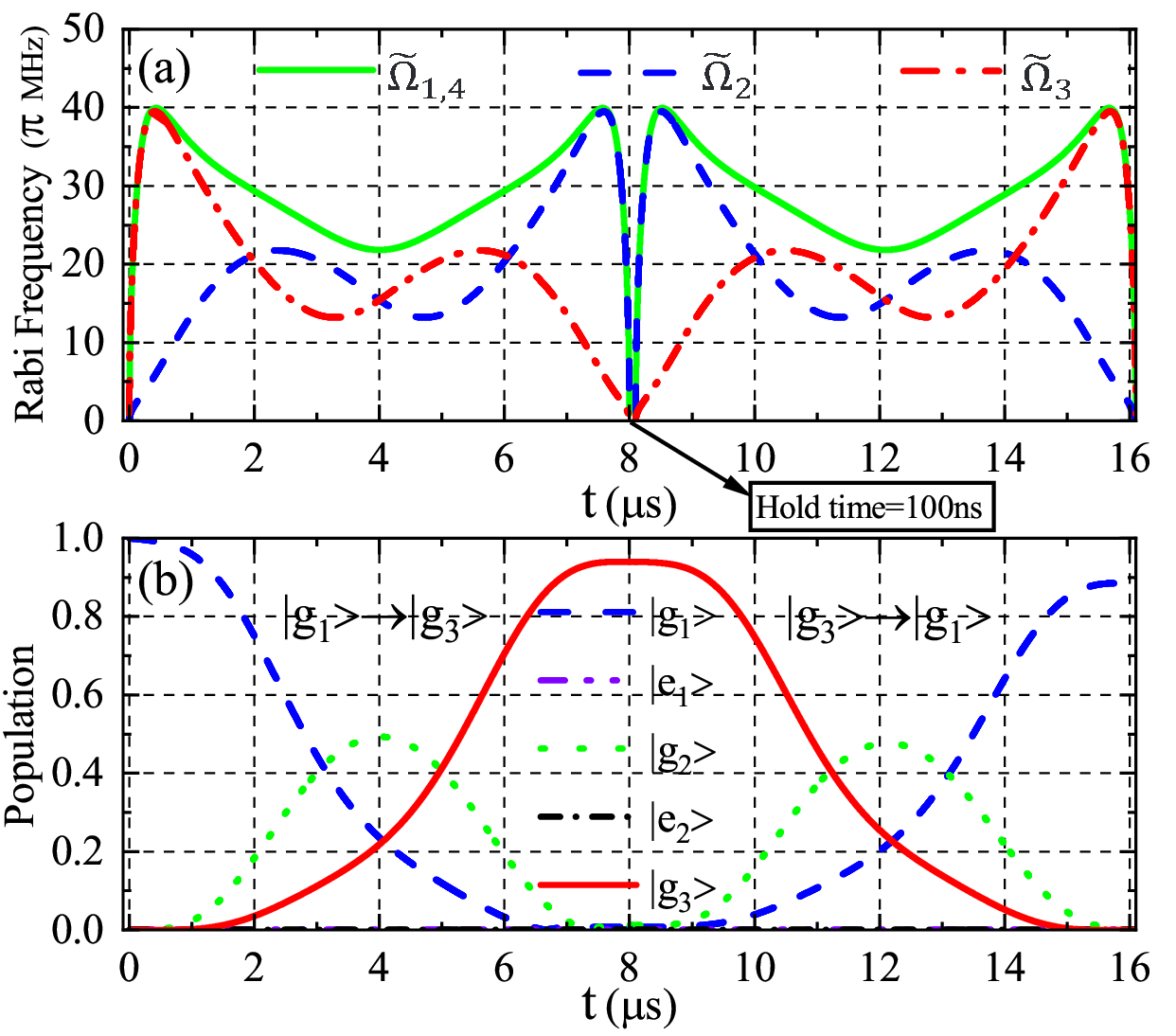}}
\caption{(a) Time sequence of the four Rabi frequencies $\tilde{\Omega}_{j} (j=1,2,3,4)$
for the creation and detection of ultracold deeply-bound molecules and
(b) the corresponding round-trip population transfer.
All other parameters are the same as those in Fig.~\ref{fig5}(b) and (c).}
\label{fig7}
\end{figure}

Finally, we discuss briefly the detection of ultracold deeply-bound molecules.
Fig.~\ref{fig7} directly gives corresponding examples of the detections following the creation results in Fig.~\ref{fig5}(b) and (c).
In order to reverse the time sequence of the four Rabi frequencies
$\tilde{\Omega}_{j} (j=1,2,3,4)$,
the boundary condition of the parameter $\vartheta(t)$ should be altered to be
$\vartheta(0)=\pi/2, \vartheta(t_f)=0$.
\noindent
The calculations in Fig.~\ref{fig7} clearly show the high efficiency of detection under our protocol. During the round-trip transfer, the pulses are smooth without any singularity, all of which can be generated with an arbitrary waveform generator in experiment \cite{PhysRevA.100.043413}.

\section{\label{sec:level4}SUMMARY}
In this paper, a theoretical method for the efficient and robust creation and detection of deeply-bound molecules based on the ``Invariant-based Inverse Engineering'' has been suggested and analyzed. The simple three- and five-level molecular systems with states chainwise coupled by optical fields are considered. The proposed method is simpler than the usual ``STA'' method, which prescribes shortcut fields between all states of the system, while the present protocols only need to modify the original fields. The key of our protocol is to first reduce the dynamics into their effective counterparts via adiabatic elimination together with the requirements of the relation among the incident pulses. These effective counterparts can then be directly compatible with different ``Invariant-based Inverse Engineering'' recipes. Thereafter, the modified fields are used to implement the target state transfer without additional coupling.
Numerical analysis suggests all modified scheme substantially improves molecular formation without strong laser pulses, and the stability against parameter variations is well preserved. In addition, all the protocols have shown good experimental feasibility. Finally, I hope that this work could provide substantial help for laboratory researchers in the preparation of gases with high-phase space density.

\section*{Acknowledgments}
I would like to thank the anonymous referees for constructive comments that are helpful for improving the quality of the work.

\bibliography{Arxiv}
\end{document}